\documentclass[reprint,amsmath,amssymb,aps]{article}
\usepackage{authblk}
\usepackage{graphicx}
\usepackage{amsmath,amssymb}
\usepackage{mathrsfs}
\usepackage{booktabs} 
\usepackage{subfigure}  
\usepackage{bbm}
\usepackage{slashed}
\usepackage{dcolumn}
\usepackage{multirow}
\usepackage{bm}
\usepackage[title]{appendix}
\usepackage{makecell}
\allowdisplaybreaks[4]
\usepackage{url}
\usepackage[colorlinks=true,citecolor=blue,urlcolor=blue,linkcolor=blue]{hyperref}
\usepackage[numbers,sort&compress]{natbib}
\usepackage[left=0.8in,right=0.8in,top=1in,bottom=1in]{geometry}
\begin{document}
	
	\title{Semileptonic decay of $\Lambda_b^0 \to \Lambda_c (2860)^+/\Lambda_c(2625)^+\ell^-\overline{\nu}_\ell$ within QCD light-cone sum rules}
	\author{Hui-Hui Duan\footnote{duanhuihui@htu.edu.cn}}
\author{Jia-Bao Feng}
\author{Feng-Mei Liu}
\author{Qin Chang\footnote{changqin@htu.edu.cn}}
\affil{School of Physics, Henan Normal University, Xinxiang 453007, Henan, People's Republic of China}
\renewcommand*{\Affilfont}{\small\it}
                 
    \date{}
\maketitle

    \begin{abstract}
    In this work, we calculate the transition form factors for the weak decays $\Lambda_b^0 \to \Lambda_c(2860)^+$ and $\Lambda_b^0 \to \Lambda_c(2625)^+$ using QCD light-cone sum rules, and compute the branching fractions of the corresponding semileptonic decays $\Lambda_b^0 \to \Lambda_c(2860)^+ \ell^- \bar{\nu}_\ell$ and $\Lambda_b^0 \to \Lambda_c(2625)^+ \ell^- \bar{\nu}_\ell$. Our predicted branching fraction for $\Lambda_b^0 \to \Lambda_c(2625)^+ \ell^- \bar{\nu}_\ell$ is consistent with experimental data and other theoretical predictions, validating the reliability of our method. On this basis, we also present the branching fraction of $\Lambda_b^0 \to \Lambda_c(2860)^+ \ell^- \bar{\nu}_\ell$. These results may serve as a theoretical reference for future experimental measurements of this decay channel.
     \end{abstract}

\section{Introduction} \label{sec:I}  
	
  In our previous work, we calculated the weak decay form factors and semileptonic decay branching fractions of $\Lambda_b\to\Lambda_c\ell\overline{\nu}_\ell$ within the framework of QCD light-cone sum rules, using the light-cone distribution amplitudes of the initial-state $\Lambda_b$ baryon \cite{Duan:2022uzm,Duan:2024lnw}. The obtained results were found to be in good agreement with experimental data and other theoretical approaches. In that work, we considered the transition from the ground-state $\Lambda_b$ baryon to the ground-state $\Lambda_c$ baryon, corresponding to the spin-parity quantum numbers $\frac{1}{2}^+\to\frac{1}{2}^+$. Subsequently, within the same framework, we computed the $\frac{1}{2}^+\to\frac{1}{2}^-$ process associated with $\Lambda_b\to\Lambda_c(2595)$ and obtained results consistent with other theoretical predictions. Therefore, in the present work, we aim to further investigate processes with higher quantum numbers for the final-state baryon, namely $\frac{1}{2}^+\to\frac{3}{2}^\pm$, corresponding to $\Lambda_b^0\to\Lambda_c(2860)^+/\Lambda_c(2625)^+$.

In this paper, we plan to compute the weak decay form factors and the corresponding semileptonic decay branching fractions for the $\Lambda_b^0\to\Lambda_c(2860)^+$ and $\Lambda_c(2625)^+$ transition processes. The decay channel $\Lambda_b^0\to\Lambda_c(2625)^+\ell^-\overline{\nu}_\ell$ was first measured by the CDF Collaboration in 2008~\cite{CDF:2008hqh}. According to the latest update from the Particle Data Group (PDG), the branching fraction is $\mathcal{B}r(\Lambda_b^0\to\Lambda_c(2625)^+\ell\overline{\nu}_\ell)=1.3^{+0.6}_{-0.5}\%$~\cite{ParticleDataGroup:2024cfk}. On the theoretical side, calculations of the $\Lambda_b^0\to\Lambda_c(2625)^+$ form factors have been performed using lattice QCD~\cite{Meinel:2021mdj,Meinel:2021rbm}, heavy quark effective theory~\cite{DiRisi:2023npw,Pervin:2005ve}, heavy quark spin symmetry~\cite{Nieves:2019kdh}, QCD light-cone sum rules~\cite{Aliev:2023tpk}, light-front quark model~\cite{Li:2022hcn}, and covariant confined quark model~\cite{Leibovich:1997az,Gutsche:2018nks}, as well as studies of new physics effects in this decay channel~\cite{Du:2022ipt,Papucci:2021pmj}. The above experimental and theoretical works provide rich material for our study of weak decays of initial spin-$\frac{1}{2}$ heavy baryons into final spin-$\frac{3}{2}$ baryons. However, these investigations have so far mainly focused on the transition from an initial state with spin-parity $\frac{1}{2}^+$ to a final state with spin-parity $\frac{3}{2}^-$, while the $\frac{1}{2}^+\to\frac{3}{2}^+$ process has received less attention.

In this work, we will employ QCD light-cone sum rules as the theoretical tool to calculate the weak decay form factors for $\Lambda_b^0\to\Lambda_c(2625)^+$ and $\Lambda_b^0\to\Lambda_c(2860)^+$, and then evaluate the semileptonic decay branching fractions of $\Lambda_b^0\to\Lambda_c(2860)^+\ell^-\overline{\nu}_\ell$ and $\Lambda_b^0\to\Lambda_c(2625)^+\ell^-\overline{\nu}_\ell$. The spin-parity quantum numbers of $\Lambda_c(2625)^+$ and $\Lambda_c(2860)^+$ are $\frac{3}{2}^-$ and $\frac{3}{2}^+$, respectively.

The remainder of this paper is organized as follows. In Sec. \ref{sec:II}, we present the basic scheme for calculating the semileptonic decay branching fractions. In Sec. \ref{sec:III}, we compute the weak decay form factors using the QCD light-cone sum rule method. Numerical analyses are given in Sec. \ref{sec:IV}. Finally, Sec. \ref{sec:V} contains our summary and conclusion.
	
   \section{Semileptonic decay of $\Lambda_b^0\to\Lambda_c(2860)^+/\Lambda_c(2625)^+\ell^-\overline{\nu}_\ell$} \label{sec:II}
		
     Phenomenologically, the observables for semileptonic decays are the decay rate or the branching fraction. Hence, our goal is to compute the branching fractions of the semileptonic processes $\Lambda_b^0\to\Lambda_c(2625)^+\ell^-\overline{\nu}_\ell$ and $\Lambda_b^0\to\Lambda_c(2860)^+\ell^-\overline{\nu}_\ell$ that can be analyzed experimentally. To this end, we first need the theoretical expressions for the decay widths of these two channels. Following the discussions in Refs.~\cite{Faessler:2009xn} and \cite{Gutsche:2017wag}, we present the formula for the decay width as follows:
	\begin{align}
	\Gamma=\frac{1}{2}\frac{G_F^2|V_{\rm CKM}|^2}{192\pi^3}\frac{M_{\Lambda_{\pm}}}{M_{\Lambda_b}^2}\int_{m_\ell^2}^{M_-^2} \frac{dq^2}{q^2}(q^2-m_\ell^2)^2 \frac{(M_{\Lambda_b}^2+M_{\Lambda_{\pm}}^2-q^2)^2-4M_{\Lambda_b}^2M_{\Lambda_\pm}^2}{2M_{\Lambda_b}M_{\Lambda_\pm}}\mathcal{H}_{\frac{1}{2}\to\frac{3}{2}}
	\end{align}
where $G_F$ is the Fermi constant, $|V_{\rm CKM}|$ is the CKM matrix element. $M_{\Lambda_b^0}$, $M_{\Lambda_-}$ and  $M_{\Lambda_+}$ are the masses of $\Lambda_b^0$, $\Lambda_c(2625)^+$ and $\Lambda_c(2860)^+$, respectively. The symbol $\mathcal{H}_{\frac{1}{2}\to\frac{3}{2}}$ contains the helicity amplitudes of the baryon with spin-$\frac{1}{2}$ decay to spin-$\frac{3}{2}$ final state, and it can be written by:
	\begin{align}
	\mathcal{H}_{\frac{1}{2}\to\frac{3}{2}}=&|H_{\frac{1}{2}1}|^2+|H_{-\frac{1}{2}-1}|^2+|H_{\frac{3}{2}1}|^2+|H_{-\frac{3}{2}-1}|^2+|H_{\frac{1}{2}0}|^2+|H_{-\frac{1}{2}0}|^2 \notag \\
	&+\frac{m_\ell^2}{2q^2}\left( 3|H_{\frac{1}{2}t}|^2+3|H_{-\frac{1}{2}t}|^2+|H_{-\frac{1}{2}1}|^2+|H_{-\frac{1}{2}-1}|^2+|H_{\frac{3}{2}1}|^2+|H_{-\frac{3}{2}-1}|^2+|H_{\frac{1}{2}0}|^2+|H_{-\frac{1}{2}0}|^2 \right)
	\end{align}
	and the helicity amplitudes $H_{\lambda \lambda_W}$ are related to the form factors defined from the hadron transition matrix elements. 
	
	For the weak decay matrix element of $\frac{1}{2}^+\to\frac{3}{2}^+$ and $\frac{1}{2}^+\to\frac{3}{2}^-$ processes, the initial baryon state to final baryon state transition  matrix element can be expressed by eight form factors respectivly, and they are
	\begin{align}
	\langle \Lambda_+(p^\prime)\vert j_\nu\vert\Lambda_b(p)\rangle&=\bar{u}^\alpha(p^\prime)\left[f_1^V(q^2)g_{\alpha\nu}+p_\alpha\left(\frac{1}{M_{\Lambda_b}}f_2^V(q^2)\gamma_\nu+\frac{1}{M_{\Lambda_b}^2}f_3^V(q^2)p_\nu^\prime+\frac{1}{M_{\Lambda_b}^2}f_4^V(q^2)q_\nu\right)\right]\gamma_5u(p) \notag \\
	&-\bar{u}^\alpha(p^\prime)\left[g_1^A(q^2)g_{\alpha\nu}+p_\alpha\left(\frac{1}{M_{\Lambda_b}}g_2^A(q^2)\gamma_\nu+\frac{1}{M_{\Lambda_b}^2}g_3^A(q^2)p_\nu^\prime+\frac{1}{M_{\Lambda_b}^2}g_4^A(q^2)q_\nu\right)\right]u(p) \label{ffp}
	\end{align}
	for $\frac{1}{2}^+\to \frac{3}{2}^+$, and
	\begin{align}
	\langle \Lambda_+(p^\prime)\vert j_\nu\vert\Lambda_b(p)\rangle&=\bar{u}^\alpha(p^\prime)\left[g_1^V(q^2)g_{\alpha\nu}+p_\alpha\left(\frac{1}{M_{\Lambda_b}}g_2^V(q^2)\gamma_\nu+\frac{1}{M_{\Lambda_b}^2}g_3^V(q^2)p_\nu^\prime+\frac{1}{M_{\Lambda_b}^2}g_4^V(q^2)q_\nu\right)\right]u(p) \notag \\
	&-\bar{u}^\alpha(p^\prime)\left[f_1^A(q^2)g_{\alpha\nu}+p_\alpha\left(\frac{1}{M_{\Lambda_b}}f_2^A(q^2)\gamma_\nu+\frac{1}{M_{\Lambda_b}^2}f_3^A(q^2)p_\nu^\prime+\frac{1}{M_{\Lambda_b}^2}f_4^A(q^2)q_\nu\right)\right]\gamma_5u(p) \label{ffn}
	\end{align}
	for $\frac{1}{2}^+\to\frac{3}{2}^-$.
	
	The detail relations between form factors and helicity amplitudes are
	\begin{align}
	H_{\frac{1}{2}t}^V&=-\sqrt{\frac{2}{3}\cdot\frac{Q_+}{q^2}}\frac{Q_-}{2M_{\Lambda_b}M_{\Lambda_+}}\left(f_1^VM_{\Lambda_b}-f_2^VM_+ +f_3^V\frac{M_+M_- -q^2}{2M_{\Lambda_b}}+f_4^V\frac{q^2}{M_{\Lambda_b}}\right) \notag \\
	H_{\frac{1}{2}0}^V&=-\sqrt{\frac{2}{3}\cdot\frac{Q_-}{q^2}}\left(f_1^V\frac{M_+M_- -q^2}{2M_{\Lambda_+}}-f_2^V\frac{Q_+M_-}{2M_{\Lambda_b}M_{\Lambda_+}}+f_3^V\frac{(M_{\Lambda_b}^2+M_{\Lambda_+}^2-q^2)^2-4M_{\Lambda_b}^2M_{\Lambda_+}^2}{2M_{\Lambda_b}M_{\Lambda_+}^2}\right) \notag \\
	H_{\frac{1}{2}1}^V&=\sqrt{\frac{Q_-}{3}}\left(f_1^V-f_2^V\frac{Q_+}{M_{\Lambda_b}M_{\Lambda_+}}\right) \notag \\
	H_{\frac{3}{2}1}^V&=-\sqrt{Q_-}f_1^V \notag \\
	H_{\frac{1}{2}t}^A&=\sqrt{\frac{2}{3}\cdot\frac{Q_-}{q^2}}\frac{Q_+}{2M_{\Lambda_b}M_{\Lambda_+}}\left(g_1^AM_{\Lambda_b}+g_2^AM_- +g_3^A\frac{M_+M_- -q^2}{2M_{\Lambda_b}}+g_4^A\frac{q^2}{M_{\Lambda_b}}\right) \notag \\
	H_{\frac{1}{2}0}^A&=\sqrt{\frac{2}{3}\cdot\frac{Q_+}{q^2}}\left(g_1^A\frac{M_+M_- -q^2}{2M_{\Lambda_+}}+g_2^A\frac{Q_-M_+}{2M_{\Lambda_b}M_{\Lambda_+}}+g_3^A\frac{(M_{\Lambda_b}^2+M_{\Lambda_+}^2-q^2)^2-4M_{\Lambda_b}^2M_{\Lambda_+}^2}{2M_{\Lambda_b}M_{\Lambda_+}^2}\right) \notag \\
	H_{\frac{1}{2}1}^A&=\sqrt{\frac{Q_+}{3}}\left(g_1^A-g_2^A\frac{Q_-}{M_{\Lambda_b}M_{\Lambda_+}}\right) \notag \\
	H_{\frac{3}{2}1}^A&=\sqrt{Q_+}g_1^A
	\end{align}
	for $\frac{1}{2}^+\to\frac{3}{2}^+$; and
	\begin{align}
	H_{\frac{1}{2}t}^V&=\sqrt{\frac{2}{3}\frac{Q_-}{q^2}}\frac{Q_+}{2M_{\Lambda_b}M_{\Lambda_-}}\left(M_{\Lambda_b} f_1^A+M_-f_2^A+\frac{M_+M_- -q^2}{2M_{\Lambda_b}}f_3^A+\frac{q^2}{M_{\Lambda_b}}f_4^A\right) \notag \\
	H_{\frac{1}{2}0}^V&=\sqrt{\frac{2}{3}\frac{Q_+}{q^2}}\left(\frac{M_+M_- -q^2}{2M_{\Lambda_-}}f_1^A+\frac{Q_-M_+}{2M_{\Lambda_b}M_{\Lambda_-}}f_2^A+\frac{(M_{\Lambda_b}^2+M_{\Lambda_-}^2-q^2)^2-4M_{\Lambda_b}^2M_{\Lambda_-}^2}{2M_{\Lambda_b}M_{\Lambda_-}^2}f_3^A\right) \notag \\
	H_{\frac{1}{2}1}^V&=\sqrt{\frac{Q_+}{3}}\left(f_1^A-\frac{Q_-}{M_{\Lambda_b}M_{\Lambda_-}}f_2^A\right) \notag \\
	H_{\frac{3}{2}1}^V&=\sqrt{Q_+}f_1^A \notag\\
	H_{\frac{1}{2}t}^A&=-\sqrt{\frac{2}{3}\frac{Q_+}{q^2}}\frac{Q_-}{2M_{\Lambda_b}M_{\Lambda_-}}\left(M_{\Lambda_b}g_1^V-M_+g_2^V+\frac{M_+M_--q^2}{2M_{\Lambda_b}}g_3^V+\frac{q^2}{M_{\Lambda_b}}g_4^V\right) \notag\\
	H_{\frac{1}{2}0}^A&=-\sqrt{\frac{2}{3}\frac{Q_-}{q^2}}\left(\frac{M_+M_--q^2}{2M_{\Lambda_-}}g_1^V-\frac{Q_+M_-}{2M_{\Lambda_b}M_{\Lambda_-}}g_2^V+\frac{(M_{\Lambda_b}^2+M_{\Lambda_-}^2-q^2)^2-4M_{\Lambda_b}^2M_{\Lambda_-}^2}{2M_{\Lambda_b}M_{\Lambda_-}^2}g_3^V\right) \notag \\
	H_{\frac{1}{2}1}^A&=\sqrt{\frac{Q_-}{q^2}}\left(g_1^V-\frac{Q_+}{M_{\Lambda_b}M_{\Lambda_-}}g_2^V\right) \notag \\
	H_{\frac{3}{2}1}^A&=-\sqrt{Q_-}g_1^V
	\end{align}
	for $\frac{1}{2}^+\to\frac{3}{2}^-$. Where $M_\pm=M_{\Lambda_b}\pm M_{\Lambda_\pm}$ and $Q_\pm=(M_{\Lambda_b}\pm M_{\Lambda_\pm})^2-q^2$. 
	\section{Theoretical framework of $\Lambda_b^0\to\Lambda_c(2860)^+/\Lambda_c(2625)^+$ transition form factors} \label{sec:III}
	
	In order to calculate the decay width and branching fractions of semileptonic decay, we need to know the momentum dependence of weak decay form factors defined in Eq.(\ref{ffp}) and (\ref{ffn}). In this section, we will calculate these form factors within QCD light-cone sum rules.
	
	Our starting point to calculate the weak decay form factors is the correlation function of hadron weak transition:
	\begin{align}
	\Pi_{\mu\nu}(p,q)=i\int d^4 x e^{ip^\prime\cdot x}\langle 0\vert \mathcal{T}\{\eta_\mu(x), j_\nu(0)\}\vert \Lambda_b(p)\rangle. \label{cor}
	\end{align}
	where $\eta_\mu(x)=[\partial_\alpha \partial_\beta(q_1(x)\mathcal{C}\gamma_5 q_2(x))]\Gamma_{\alpha\beta\mu}c(x)$ is the interpolating current of the final baryon $\Lambda_c(2860)^+$ within spin-parity quantum number $J^P=\frac{3}{2}^+$, $\Gamma_{\alpha\beta\mu}=(g_{\alpha\mu}g_{\beta\tau}+g_{\alpha\tau}g_{\beta\mu}-\frac{1}{2}g_{\alpha\beta}g_{\mu\tau})\gamma^\tau\gamma_5$ \cite{Wang:2017vtv}. $j_\nu(0)=\bar{c}(0)\gamma_\nu(1-\gamma_5)b(0)$ is the standard $V - A$ weak decay current of bottom to charm quark, $p$ is the momentum of initial $\Lambda_b^0$ baryon, $p^\prime$ is the momentum of final charm baryon, $q$ is the momentum transfer from the weak decay of initial baryon to final baryon and fulfill the relation $q^2=(p-p^\prime)^2$. Actually, $q^2$ represent the momentum of final leptons in the semileptonic decay.
	
	The next task is the calculation on both hadronic and QCD level of correlation (\ref{cor}).
	
	On the hadronic level, the final form of correlation function can be expressed as:
	\begin{align}
	\Pi_{\mu\nu}(p,q)=-\frac{\lambda_+}{M_{\Lambda_+}^2-p^{\prime 2}}&\left\{(M_{\Lambda_+}-M_{\Lambda_b})f_1^Vg_{\mu\nu}\gamma_5-f_1^Vg_{\mu\nu}\slashed{q}\gamma_5 \right. \notag \\
	&+p_\mu\left[\left(\frac{2}{M_{\Lambda_b}}f_2^V-\frac{M_{\Lambda_b}-M_{\Lambda_+}}{M_{\Lambda_b}^2}f_3^V\right)p_\nu\gamma_5+\frac{M_{\Lambda_b}+M_{\Lambda_+}}{M_{\Lambda_b}}f_2^V\gamma_\nu\gamma_5 \right. \notag \\ 
	&+\left(-\frac{2}{M_{\Lambda_b}}f_2^V+\frac{M_{\Lambda_b}-M_{\Lambda_+}}{M_{\Lambda_b}^2}f_3^V-\frac{M_{\Lambda_b}-M_{\Lambda_+}}{M_{\Lambda_b}^2}f_4^V)\right)q_\nu\gamma_5  \notag \\
	&+\left.\frac{1}{M_{\Lambda_b}}f_2^V\gamma_\nu\slashed{q}\gamma_5-\frac{1}{M_{\Lambda_b}^2}f_3^Vp_\nu\slashed{q}\gamma_5+\frac{1}{M_{\Lambda_b}^2}\left(f_3^V-f_4^V\right)q_\nu\slashed{q}\gamma_5\right]  \notag \\
	&-(M_{\Lambda_+}+M_{\Lambda_b})g_1^Ag_{\mu\nu}+g_1^Ag_{\mu\nu}\slashed{q}  \notag \\
	&+p_\mu\left[-\left(\frac{2}{M_{\Lambda_b}}g_2^A+\frac{M_{\Lambda_b}+M_{\Lambda_+}}{M_{\Lambda_b}^2}g_3^A\right)p_\nu+\frac{M_{\Lambda_b}-M_{\Lambda_+}}{M_{\Lambda_b}}g_2^A\gamma_\nu \right. \notag \\ 
	&+\left(\frac{2}{M_{\Lambda_b}}g_2^A+\frac{M_{\Lambda_b}+M_{\Lambda_+}}{M_{\Lambda_b}^2}g_3^A-\frac{M_{\Lambda_b}+M_{\Lambda_+}}{M_{\Lambda_b}^2}g_4^A)\right)q_\nu  \notag \\
	&-\left.\frac{1}{M_{\Lambda_b}}g_2^A\gamma_\nu\slashed{q}+\frac{1}{M_{\Lambda_b}^2}g_3^Ap_\nu\slashed{q}-\frac{1}{M_{\Lambda_b}^2}\left(g_3^A-g_4^A\right)q_\nu\slashed{q}\right] \left.\right\}u(p)	\notag \\
	-\frac{\lambda_-}{M_{\Lambda_-}^2-p^{\prime 2}}&\left\{(M_{\Lambda_-}+M_{\Lambda_b})g_1^Vg_{\mu\nu}\gamma_5+g_1^Vg_{\mu\nu}\slashed{q}\gamma_5 \right. \notag \\
	&+p_\mu\left[\left(\frac{2}{M_{\Lambda_b}}g_2^V+\frac{M_{\Lambda_b}+M_{\Lambda_-}}{M_{\Lambda_b}^2}g_3^V\right)p_\nu\gamma_5+\frac{M_{\Lambda_b}-M_{\Lambda_-}}{M_{\Lambda_b}}g_2^V\gamma_\nu\gamma_5 \right. \notag \\ 
	&+\left(-\frac{2}{M_{\Lambda_b}}g_2^V-\frac{M_{\Lambda_b}+M_{\Lambda_-}}{M_{\Lambda_b}^2}g_3^V+\frac{M_{\Lambda_b}+M_{\Lambda_-}}{M_{\Lambda_b}^2}g_4^V)\right)q_\nu\gamma_5  \notag \\
	&+\left.\frac{1}{M_{\Lambda_b}}g_2^V\gamma_\nu\slashed{q}\gamma_5+\frac{1}{M_{\Lambda_b}^2}g_3^Vp_\nu\slashed{q}\gamma_5-\frac{1}{M_{\Lambda_b}^2}\left(g_3^V-g_4^V\right)q_\nu\slashed{q}\gamma_5\right] \notag \\
	&+(M_{\Lambda_b}-M_{\Lambda_-})f_1^Ag_{\mu\nu}-f_1^Ag_{\mu\nu}\slashed{q}  \notag \\
	&+p_\mu\left[\left(-\frac{2}{M_{\Lambda_b}}f_2^A+\frac{M_{\Lambda_b}-M_{\Lambda_-}}{M_{\Lambda_b}^2}f_3^A\right)p_\nu+\frac{M_{\Lambda_b}+M_{\Lambda_-}}{M_{\Lambda_b}}f_2^A\gamma_\nu \right. \notag \\ 
	&+\left(\frac{2}{M_{\Lambda_b}}f_2^A-\frac{M_{\Lambda_b}-M_{\Lambda_-}}{M_{\Lambda_b}^2}f_3^A+\frac{M_{\Lambda_b}-M_{\Lambda_-}}{M_{\Lambda_b}^2}f_4^A)\right)q_\nu  \notag \\
	&-\left.\frac{1}{M_{\Lambda_b}}f_2^A\gamma_\nu\slashed{q}-\frac{1}{M_{\Lambda_b}^2}f_3^Ap_\nu\slashed{q}+\frac{1}{M_{\Lambda_b}^2}\left(f_3^A-f_4^A\right)q_\nu\slashed{q}\right] \left.\right\}u(p) \notag \\
	&+\text{higher states and continuum}.
	\end{align}
The phrase `higher states and continuum' denotes the higher excited states and continuum of the final charm baryon; these contributions will be separated by introducing a threshold parameter $s_0$ in the dispersion integral, and a detailed discussion will be given in the numerical section.
	
	It should be noted here that when calculating the form factors for the final-state baryon with spin $\frac{3}{2}^+$ in weak decays, the Lorentz structures required are different from those for the final-state baryon with spin $\frac{1}{2}$ \cite{Duan:2024lnw}. However, in the calculation of correlation functions, both types of final-state spin structures appear simultaneously. At the hadronic level of the correlation function, we need to discard the parts associated with the final-state spin-$\frac{1}{2}$ baryon and retain only those associated with the final-state spin-$\frac{3}{2}$ baryon. Similarly, at the QCD level, we also need to discard the parts related to the contributions of the final-state spin-$\frac{1}{2}$ baryon. Specifically, in the contribution of the final-state baryon with spin $\frac{1}{2}$, the Lorentz structure contains only one subscript $\mu$ \cite{Duan:2024lnw}, whereas in the contribution of the final-state baryon with spin $\frac{3}{2}$, the Lorentz structure contains both subscripts $\mu$ and $\nu$. For a detailed discussion on this aspect, see Ref.~\cite{Aliev:2023tpk,Huang:2024oik}.
	
	At the QCD, through the Wick contraction of the hadronic interpolating currents and the weak decay currents in the correlation function, we finally obtain
\begin{align}
\Pi_{\mu\nu}(p,q)=i\int d^4 x e^{ip^\prime \cdot x} (\mathcal{C}\gamma_5)_{\phi\eta}(\Gamma_{\alpha\beta\mu} S(x)\gamma_\nu(1-\gamma_5))_{\kappa\gamma}\epsilon_{abc}\partial_\alpha\partial_\beta\langle 0|q_{1\phi}^{aT}(x)q_{2\eta}^b(x)b_\gamma^c(0)|\Lambda_b^0(p)\rangle,
\end{align}
where $S(x)$ is the free charm quark propagator. $\langle 0|q_{1\phi}^{aT}(x)q_{2\eta}^b(x)b_\gamma^c(0)|\Lambda_b^0(p)\rangle$ is the light-cone distribution amplitude of the $\Lambda_b^0$ baryon, which has been discussed in our previous papers \cite{Duan:2022uzm,Duan:2024lnw}. Specifically, it can be expressed in terms of light-cone wave functions with different twists:
\begin{align}  
		\epsilon_{abc}\langle 0| & q^{aT}_{1\phi} (t_1n)q^b_{2\eta}(t_2n)b^c_{\gamma}| \Lambda_b^0(p)\rangle \notag \\ &= \frac{1}{8}v_+f_{\Lambda_b}^{(1)}\Psi^n(t_1,t_2)(\slashed{\bar{n}}\gamma_5C^T)_{\eta\phi}u_{\Lambda_b\gamma}(v) \notag \\
		&+\frac{1}{4}f_{\Lambda_b}^{(2)}\Psi^{\mathbbm{1}}(t_1,t_2)(\gamma_5C^T)_{\eta\phi}u_{\Lambda_b\gamma}(v) \notag \\  &
		 -\frac{1}{8}f_{\Lambda_b}^{(2)}\Psi^{n\bar{n}}(t_1,t_2)(i\sigma_{n\bar{n}}\gamma_5C^T)_{\eta\phi}u_{\Lambda_b\gamma}(v) \notag \\
		 &+ \frac{1}{8}\frac{1}{v_+}f_{\Lambda_b}^{(1)}\Psi^{\bar{n}}(t_1,t_2)(\slashed{n}\gamma_5C^T)_{\eta\phi}u_{\Lambda_b\gamma}(v).  
	\end{align}  
$t_1$ and $t_2$ represent the positions of the light quarks, $f_{\Lambda_b}^{(1)}$ and $f_{\Lambda_b}^{(2)}$ are the decay constants of $\Lambda_b^0$ baryon. The other symbols are the same as in Ref. \cite{Duan:2024lnw}. Each light-cone distribution amplitude $\{\psi^n, \psi^{n\bar{n}}, \psi^{\mathbbm{1}}, \psi^{\bar{n}}\}$ possesses a distinct twist (dimension minus spin), which we denote using the subscript numbers $\{\psi_2, \psi_3^\sigma, \psi_3^s, \psi_4\}$. And
	\begin{gather}  
		\Psi_i(t_1,t_2)=\int_{0}^{\infty}\omega d\omega\int_{0}^{1}due^{-i\omega(t_1u+t_2(1-u))}\tilde{\psi}_i(\omega,u).  
	\end{gather}  	
$\omega$ and $u$ represent the energy of the diquark and the fraction of each light quark, respectively. $\omega=\sigma M_{\Lambda_b}$, $u=0\sim 1$. Various representations of the light-cone wave functions for the $\Lambda_b^0$ baryon exist in the literature. In this work, we adopt the representation given in Ref. \cite{Bell:2013tfa}:
		\begin{align}
			\psi_2(\omega, u)=&\frac{\omega^2 u (1-u)}{\omega_0^4} e^{-\omega/\omega_0}, \notag \\
			\psi_3^s(\omega, u)=&\frac{\omega}{2\omega_0^3} e^{-\omega/\omega_0},  \notag \\
			\psi_3^\sigma(\omega, u)=&\frac{\omega(2u-1)}{2\omega_0^3}e^{-\omega/\omega_0}, \notag \\
			\psi_4(\omega, u)=&\frac{1}{\omega_0^2} e^{-\omega/\omega_0},
		\end{align}
		 where $\omega_0 \simeq 0.4~\rm{GeV}$. Finally, we obtain the correlation function on the QCD level	
	\begin{align}
	\Pi_{\mu\nu}(p,q)=&\int_0^1du\int_0^\infty d\omega\frac{f_{\Lambda_b}^{(1)}\omega^3\psi_3^s(\omega,u)}{(p^\prime-\omega v)^2-m_c^2}\left\{\frac{1}{M_{\Lambda_b}^2}\left[2\overline{\sigma}M_{\Lambda_b}-\frac{2(M_{\Lambda_b}^2-M_{\Lambda_+}^2+q^2)}{M_{\Lambda_b}}\right]p_\mu\gamma_\nu\gamma_5 \right. \notag \\
	&+\frac{4}{M_{\Lambda_b}^2}p_\mu p_\nu\slashed{q}\gamma_5+\frac{4}{M_{\Lambda_b}}p_\mu q_\nu\gamma_5-\frac{2}{M_{\Lambda_b}}p_\mu\gamma_\nu\slashed{q}\gamma_5-\frac{4m_c}{M_{\Lambda_b}^2}p_\mu p_\nu\gamma_5 \notag \\
	&-\frac{1}{M_{\Lambda_b}^2}\left[2\overline{\sigma}M_{\Lambda_b}-\frac{2(M_{\Lambda_b}^2-M_{\Lambda_+}+q^2)}{M_{\Lambda_b}}+2m_c\right]p_\mu\gamma_\nu-\frac{4}{M_{\Lambda_b}}p_\mu p_\nu\slashed{q} \notag \\
	&\left. +\frac{4}{M_{\Lambda_b}}p_\mu q_\nu-\frac{2}{M_{\Lambda_b}}p_\mu\gamma_\nu\slashed{q}+\frac{4m_c}{M_{\Lambda_b}^2}p_\mu p_\nu\right\}u(p),
		\end{align}
where $\overline{\sigma}=1-\sigma$.

	Consequently, using the quark-hadron duality relation, together with the dispersion relation and the standard procedure of QCD light-cone sum rules, we can extract the weak decay form factors for the transition processes $\Lambda_b^0\to\Lambda_c(2860)^+$ and $\Lambda_b^0\to\Lambda_c(2625)^+$. In order to suppress the contributions from higher excited states and the continuum of the interpolating hadrons in the sum rule, as well as the contributions from higher-twist terms in the light-cone distribution amplitudes of the $\Lambda_b^0$ baryon, we perform a Borel transformation on both the hadronic and the QCD levels of the correlation function, and introduce a Borel parameter. After mathching the two side of correlation function on both hadronic and QCD level by the same Lorentz structure, and introducing a cutoff threshold $s_0$ to isolate the lower state contribution. All these procedures are the same as those in our previous work \cite{Duan:2024lnw}. Therefore, we can obtain the relations of form factors
	\begin{align}
	f_1^V(q^2)&=g_1^A(q^2)=f_1^A(q^2)=g_1^V(q^2)=0; \notag \\
	f_3^V(q^2)&=f_4^V(q^2)=g_3^A(q^2)=g_4^A(q^2);  \notag \\
	f_3^A(q^2)&=f_4^A(q^2)=g_3^V(q^2)=g_4^V(q^2);  \notag \\
	\end{align}
	
	The explicit expressions of the form factors are
			\begin{align}
		f_2^V(q^2)=&\frac{f_{\Lambda_b}}{f_{\Lambda_+}}M_{\Lambda_b}^4\int_0^1 du\int_0^{\sigma_0}d\sigma\frac{\sigma^3}{1-\sigma}\psi_3^s(\omega, u)e^{(M_{\Lambda_+}^2-s)/M_B^2} \times \notag \\ & \times \frac{2}{M_{\Lambda_-}+M_{\Lambda_+}}\left\{(1-\sigma)M_{\Lambda_b}-M_{\Lambda_-}+\frac{M_{\Lambda_+}^2-q^2}{M_{\Lambda_b}}+m_c\right\}, \label{ff f2v} \\
		f_3^V(q^2)=&-\frac{f_{\Lambda_b}}{f_{\Lambda_+}}M_{\Lambda_b}^4\int_0^1 du\int_0^{\sigma_0}d\sigma\frac{\sigma^3}{1-\sigma}\psi_3^s(\omega, u)e^{(M_{\Lambda_+}^2-s)/M_B^2} \times \notag \\ & \times \frac{4}{M_{\Lambda_-}+M_{\Lambda_+}}\left(M_{\Lambda_-}+m_c\right), \\
		g_2^A(q^2)=&\frac{f_{\Lambda_b}}{f_{\Lambda_+}}M_{\Lambda_b}^4\int_0^1 du\int_0^{\sigma_0}d\sigma\frac{\sigma^3}{1-\sigma}\psi_3^s(\omega, u)e^{(M_{\Lambda_+}^2-s)/M_B^2} \times \notag \\ & \times \frac{2}{M_{\Lambda_-}+M_{\Lambda_+}}\left\{(1-\sigma)M_{\Lambda_b}+M_{\Lambda_-}+\frac{M_{\Lambda_+}^2-q^2}{M_{\Lambda_b}}+m_c\right\}, \\
		f_2^A(q^2)=&-\frac{f_{\Lambda_b}}{f_{\Lambda_+}}M_{\Lambda_b}^4\int_0^1 du\int_0^{\sigma_0}d\sigma\frac{\sigma^3}{1-\sigma}\psi_3^s(\omega, u)e^{(M_{\Lambda_+}^2-s)/M_B^2} \times \notag \\ & \times \frac{2}{M_{\Lambda_-}+M_{\Lambda_+}}\left\{(1-\sigma)M_{\Lambda_b}-M_{\Lambda_-}+\frac{M_{\Lambda_+}^2-q^2}{M_{\Lambda_b}}+m_c\right\},  \\
		f_3^A(q^2)=&\frac{f_{\Lambda_b}}{f_{\Lambda_+}}M_{\Lambda_b}^4\int_0^1 du\int_0^{\sigma_0}d\sigma\frac{\sigma^3}{1-\sigma}\psi_3^s(\omega, u)e^{(M_{\Lambda_+}^2-s)/M_B^2} \times \notag \\ & \times \frac{4}{M_{\Lambda_-}+M_{\Lambda_+}}\left(M_{\Lambda_+}-m_c\right), \\
		g_2^A(q^2)=&-\frac{f_{\Lambda_b}}{f_{\Lambda_+}}M_{\Lambda_b}^4\int_0^1 du\int_0^{\sigma_0}d\sigma\frac{\sigma^3}{1-\sigma}\psi_3^s(\omega, u)e^{(M_{\Lambda_+}^2-s)/M_B^2} \times \notag \\ & \times \frac{2}{M_{\Lambda_-}+M_{\Lambda_+}}\left\{(1-\sigma)M_{\Lambda_b}+M_{\Lambda_+}+\frac{M_{\Lambda_+}^2-q^2}{M_{\Lambda_b}}-m_c\right\} \label{ff g2a}
		\end{align}
		where
		\begin{equation}
		s=\sigma M_{\Lambda_b}^2+\frac{m_c^2-\sigma q^2}{1-\sigma}
		\end{equation}
		
		\begin{equation}
		\sigma_0=\frac{(s_0+M_{\Lambda_b}^2-q^2)-\sqrt{(s_0+M_{\Lambda_b}^2-q^2)^2-4M_{\Lambda_b}^2(s_0-m_c^2)}}{2M_{\Lambda_b}^2}
		\end{equation}
		
	In the next step, we will analyze the numerical behavior of these form factors and use them to compute the semileptonic decay branching fractions.	 
		
	
	\section{Numerical analysis} \label{sec:IV}
	
	To analyze the numerical behavior of the form factors and apply them to the calculation of the branching fractions for the semileptonic decays $\Lambda_b^0\to\Lambda_c(2860)^+/\Lambda_c(2625)^+\ell^-\overline{\nu}_\ell$, we first need to make appropriate choices for some basic parameters used in the calculation. The fundamental mass parameters are taken from the world averages given by the PDG~\cite{ParticleDataGroup:2024cfk} and are listed in Table \ref{tab mass}. For the charm quark, we adopt the pole mass. In addition, the baryon decay constants $f_{\Lambda_b}^{(1)}$ of $\Lambda_b^0$, $f_{\Lambda_+}$ of $\Lambda_c(2860)^+$ and $f_{\Lambda_-}$ of $\Lambda_c(2625)^+$ are defined and calculated within QCD sum rule method by other authors, these parameters are $f_{\Lambda_b}^{(1)}\approx f_{\Lambda_b}^{(2)}=0.03~{\rm GeV^3}$ \cite{Ball:2008fw}, $f_{\Lambda_+}=0.84~{\rm GeV^5}$ \cite{Wang:2017vtv}, and $f_{\Lambda_-}=0.041~{\rm GeV^4}$ \cite{Wang:2015kua}.
	
	    \begin{table}[h] 
	\centering  
	\caption{Mass parameters of baryons and leptons~\cite{ParticleDataGroup:2024cfk}.}  
	\label{tab mass}
		\begin{tabular}{cc}  \hline
		Parameters           &         Values              \\  
		\midrule  
		$M_{\Lambda_b^0}$    &        $5.6196$ GeV         \\  
		$M_{\Lambda_+}$      &        $2.856$ GeV        \\ 
		$M_{\Lambda_-}$    &        $2.628$ GeV        \\  
		$m_c$                &        $(1.67\pm0.07)$ GeV  \\  
		$m_e$                &        $0.51$ MeV          \\  
		$m_\mu$              &       $105.658$ MeV        \\  
		$m_\tau$             &       $1.77686$ GeV        \\  \hline
	\end{tabular}  
    \end{table}

Furthermore, in the calculation of the form factors using QCD light-cone sum rules, we introduce a threshold parameter $s_0$ to separate the contributions of the ground state from those of excited states and the continuum in the dispersion relation. To suppress the contributions of higher-twist light-cone distribution amplitudes of the $\Lambda_b^0$ baryon and higher and continuum states of final states, we perform a Borel transformation on both the hadronic and the QCD representations of the correlation function and introduce the Borel parameter $M_B$.

In this work, the only contributing $\Lambda_b^0$ baryon light-cone distribution amplitudes are those of twist-3. Therefore, when analyzing the allowed range of the Borel parameter, we only need to suppress the contributions from excited states and the continuum in the hadronic side. To ensure that the contribution of the final-state charmed baryon $\Lambda_c(2860)^+$ or $\Lambda_c(2625)^+$ exceeds 70\%, we take the Borel parameter to be $M_B^2 = (3\pm 0.1)$ $\text{GeV}^2$. The choice of the threshold parameter $s_0$ is constrained by the experimental measurements of the masses of $\Lambda_c(2860)^+$ and by the QCD sum rule calculation of these masses. In this paper, we adopt the values given in Ref.~\cite{Wang:2017vtv}, $s_0 = (M_{\Lambda_+} + \Delta)^2$, with $\Delta = (0.5 \pm 0.1)$ $\text{GeV}$.

	\subsection{Form factors} 
	
	Based on the input parameters described above and the form factor expressions given in Eq. (\ref{ff f2v}) $\sim$ (\ref{ff g2a}), we can plot the form factors as functions of the squared momentum transfer $q^2$. However, when applying the form factors to compute semileptonic decay branching fractions, the required $q^2$ dependence for the decays $\Lambda_b^0\to\Lambda_c(2860)^+\ell^-\overline{\nu}_\ell$ and $\Lambda_b^0\to\Lambda_c(2625)^+\ell^-\overline{\nu}_\ell$ lies in the physical regions $m_\ell^2 \le q^2 \le (M_{\Lambda_b^0}-M_{\Lambda_+})^2$ and $m_\ell^2 \le q^2 \le (M_{\Lambda_b^0}-M_{\Lambda_+})^2$, respectively. The form factors obtained from QCD light-cone sum rules are reliable only in the vicinity of $q^2\simeq 0~\text{GeV}^2$. Specifically, for $f_2^V(q^2)$ we can provide the $q^2$ dependence in the range $-15 \le q^2 \le 5~\text{GeV}^2$; for other form factors we can provide it in the range $-5 \le q^2 \le 2.5~\text{GeV}^2$. To obtain the form factors over the entire physical $q^2$ region, an analytic continuation is required. We adopt the well-established and widely used `$z$-series' expansion method~\cite{Bourrely:2008za}:
	\begin{equation}  
		f_i(q^2)/g_i(q^2)=\frac{1}{1-q^2/M_{B_c}^2} \left\{a_0+a_1 z(q^2, t_0)+a_2 z(q^2, t_0)^2\right\}.  \label{eqfiting}
	\end{equation}
    	Where, $a_0$, $a_1$ and $a_2$ are fitting parameters, which along with $f_i(0)/g_i(0)$, are listed in Table \ref{table 2860} and \ref{table 2625}. $M_{B_c} = 6.27447 \text{ GeV}$ represents the mass of the $B_c$ meson. The function $z(q^2, t_0)$ is defined by:  	
	\begin{equation}  
		z(q^2, t_0)=\frac{\sqrt{t_+-q^2}-\sqrt{t_+-t_0}}{\sqrt{t_+-q^2}+\sqrt{t_+-t_0}},   
	\end{equation}  
		where $t_+$ and $t_0$ are given by:  	
	\begin{align}  
		t_+ &= \left(M_{\Lambda_b} + M_{\Lambda_\pm}\right)^2, \notag  \\  
		t_0 &= \left(M_{\Lambda_b} + M_{\Lambda_\pm}\right)\cdot \left(\sqrt{M_{\Lambda_b}} - \sqrt{M_{\Lambda_\pm}}\right)^2.  
	\end{align}
	
	\begin{table}[h]
	\centering
	\caption{Form factors for the transition $\Lambda_b^0\to\Lambda_c(2860)^+$ at $q^2=0$ GeV and the fitting parameters in Eq. (\ref{eqfiting}).}
	\begin{tabular}{ccccc}\hline
	$f_i$ & $f_i(0)$ & $a_0$ & $a_1$ & $a_2$\\ \hline 
	$f_2^V$ & $0.12\pm0.03$ & $0.12\pm0.03$ & $0.50\pm0.22$  & $-10.32\pm2.81$ \\
	$f_3^V$ & $-0.55\pm0.11$ &$-0.76\pm0.15$  & $16.94^{+2.88}_{-3.02}$ & $-154.62^{+26.09}_{-24.94}$  \\
	$g_2^A$ & $0.67\pm0.14$ &  $0.86\pm0.17$ & $-14.83\pm2.48$ & $116.53^{+16.64}_{-17.52}$\\ \hline
	\end{tabular} \label{table 2860}
	\end{table}  
	
\begin{figure}
	\centering
	\includegraphics[width=0.3\textwidth]{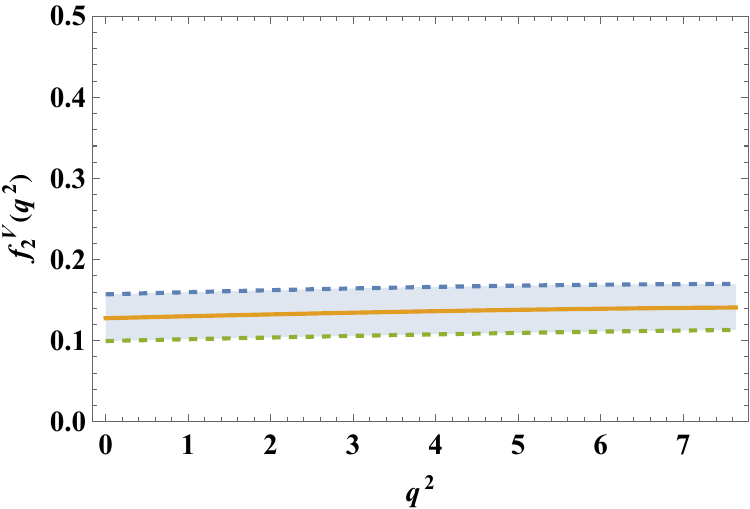}
	\includegraphics[width=0.3\textwidth]{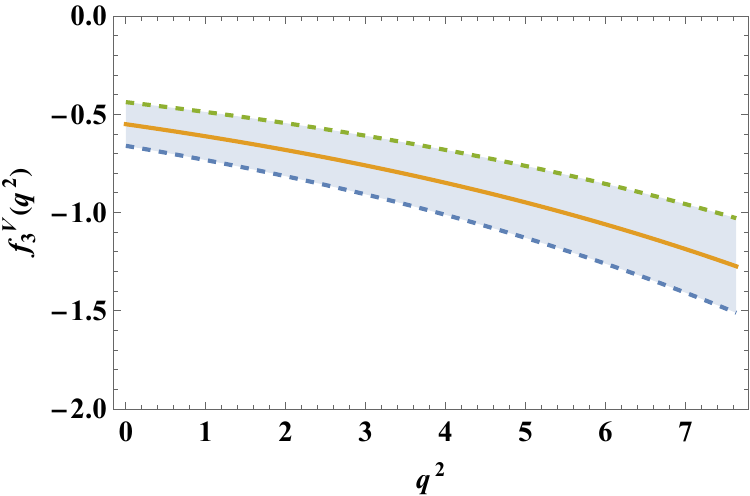}
	\includegraphics[width=0.3\textwidth]{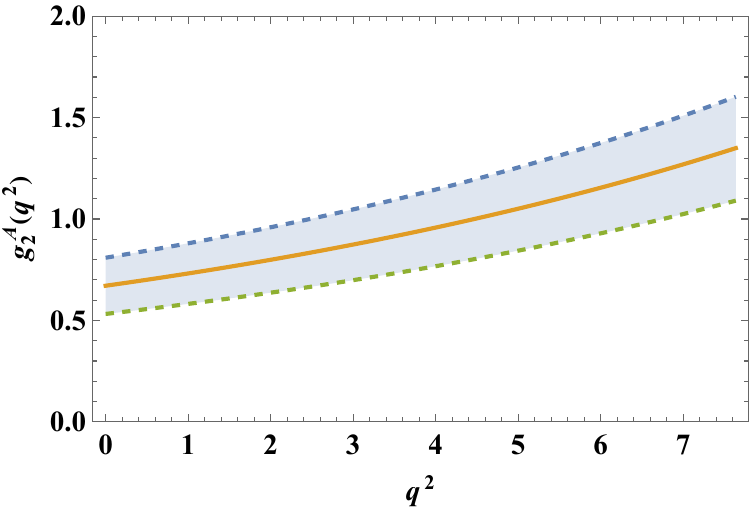}
	\caption{Form factors for the transition $\Lambda_b^0\to\Lambda_c(2860)^+$: $f_2^V$, $f_3^V$ and $g_2^A$.} \label{ff 2860}
	\end{figure}

\begin{figure}
	\centering
	\includegraphics[width=0.3\textwidth]{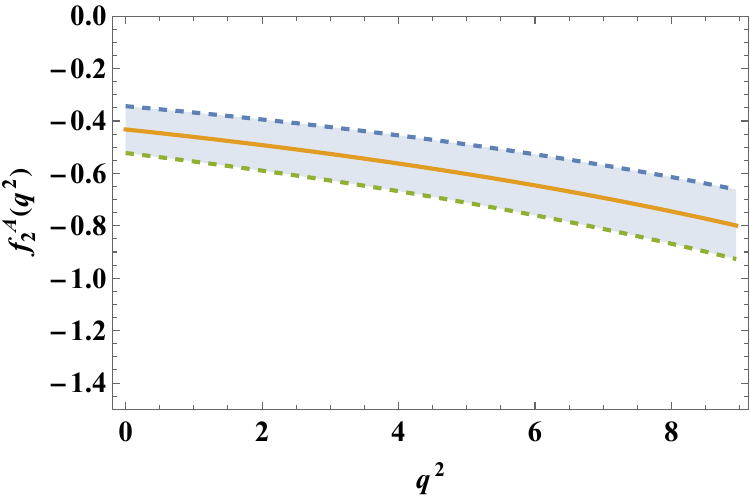}
	\includegraphics[width=0.3\textwidth]{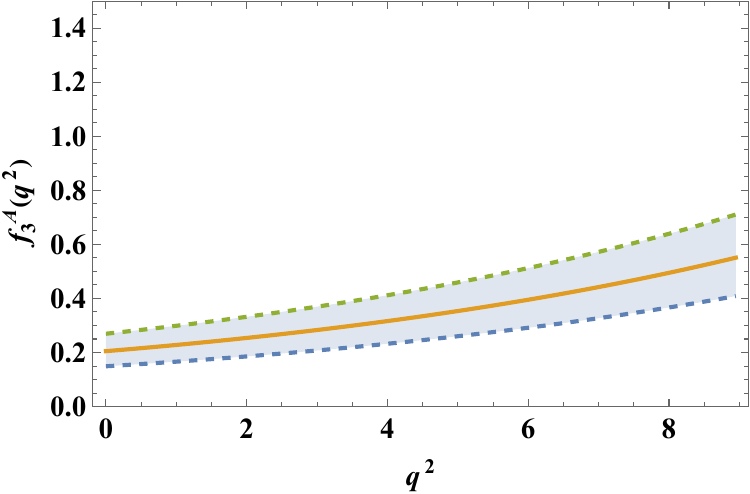}
	\includegraphics[width=0.3\textwidth]{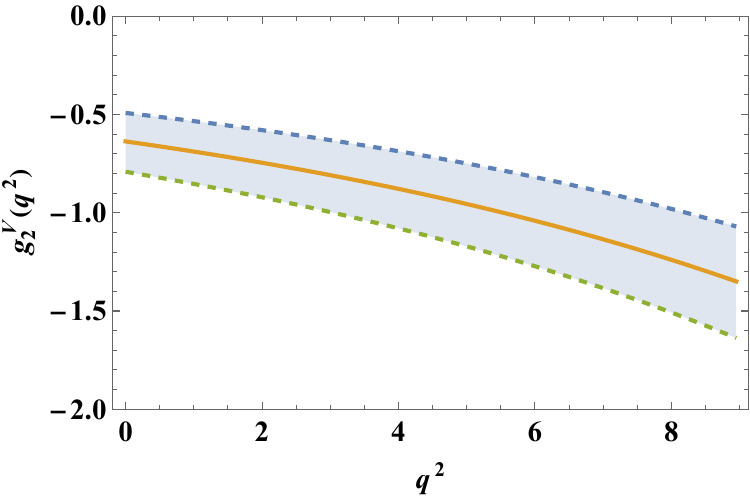}
	\caption{Form factors for the transition $\Lambda_b^0\to\Lambda_c(2625)^+$ transition: $f_2^A$, $f_3^A$ and $g_2^V$.} \label{ff 2625}
	\end{figure}

By numerically fitting the $q^2$ dependence of the form factors obtained from QCD light-cone sum rules, we obtain the fit parameters listed in Table \ref{table 2860} and \ref{table 2625}. Using these fit parameters, we can determine the $q^2$ dependence of the form factors in the whole physical region, as shown in Figs. \ref{ff 2860} and \ref{ff 2625}.

	\begin{table}
	\centering
	\caption{Form factors for the transition $\Lambda_b^0\to\Lambda_c(2625)^+$ at $q^2=0$ GeV and the fitting parameters in Eq. (\ref{eqfiting}).}
	\begin{tabular}{ccccc}\hline
	$f_i$ & $f_i(0)$ & $a_0$ & $a_1$ & $a_2$ \\ \hline 
	$f_2^A$ & $-0.43\pm0.09$ &$-0.52\pm0.10$  & $5.24^{+0.26}_{-0.49}$  & $-21.22^{+3.11}_{-1.14}$ \\
	$f_3^A$ & $0.21\pm0.06$ & $0.30^{+0.09}_{-0.08}$ & $-6.25^{+1.53}_{-1.67}$  & $51.33^{+12.87}_{-11.96}$ \\
	$g_2^V$ & $-0.64\pm0.15$ &$-0.82\pm0.18$  & $11.50^{+1.92}_{-2.02}$ &$-72.55^{+10.83}_{-9.76}$ \\ \hline
	\end{tabular} \label{table 2625}
	\end{table}

   \subsection{Branching fractions of $\Lambda_b^0 \to \Lambda_c(2860)^+/\Lambda_c(2625)^+\ell^-\overline{\nu}_\ell$}
	
\begin{figure}
	 \centering
	 \includegraphics[width=0.3\textwidth]{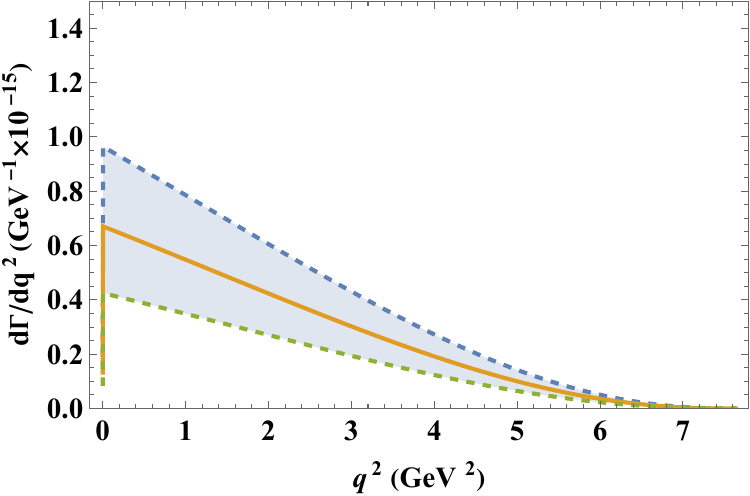}
	 \includegraphics[width=0.3\textwidth]{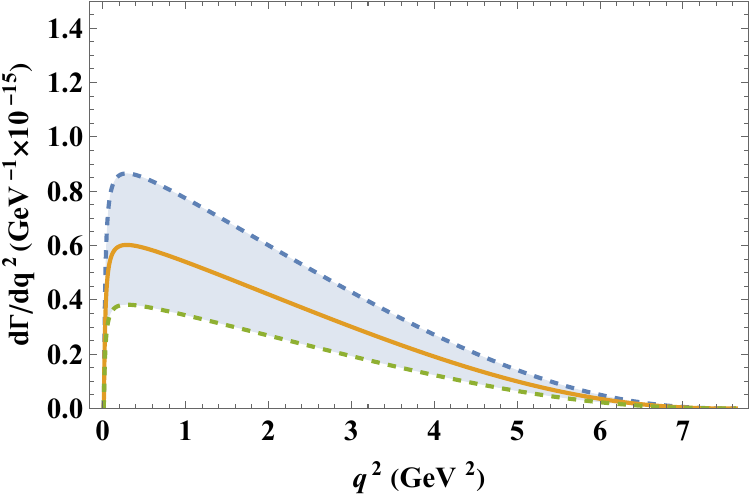}
	 \includegraphics[width=0.3\textwidth]{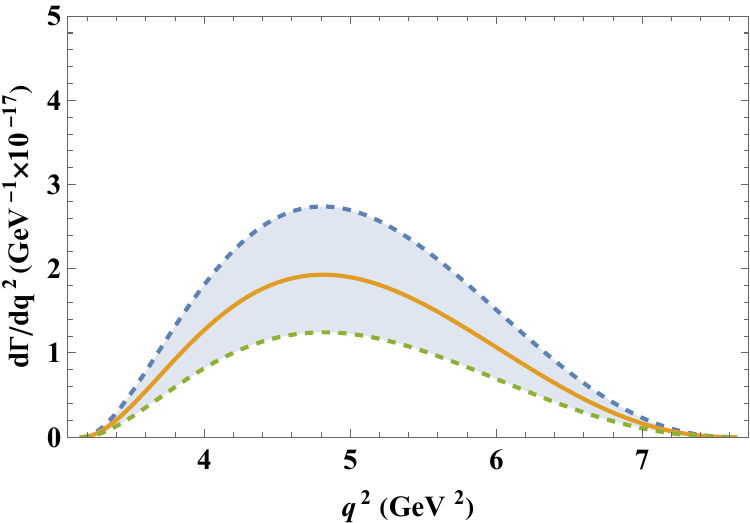}
	 \caption{Differential decay width of $\Lambda_b^0\to\Lambda_c(2860)^+\ell^-\overline{\nu}_\ell$ with $\ell=e$ (left), $\mu$ (middle) and $\tau$ (right).} \label{decay width 2860}
	 \end{figure}

\begin{figure}
	 \centering
	 \includegraphics[width=0.3\textwidth]{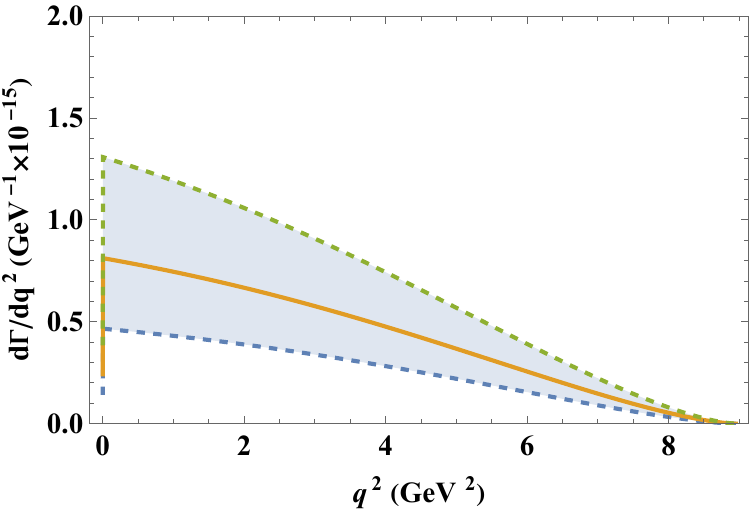}
	 \includegraphics[width=0.3\textwidth]{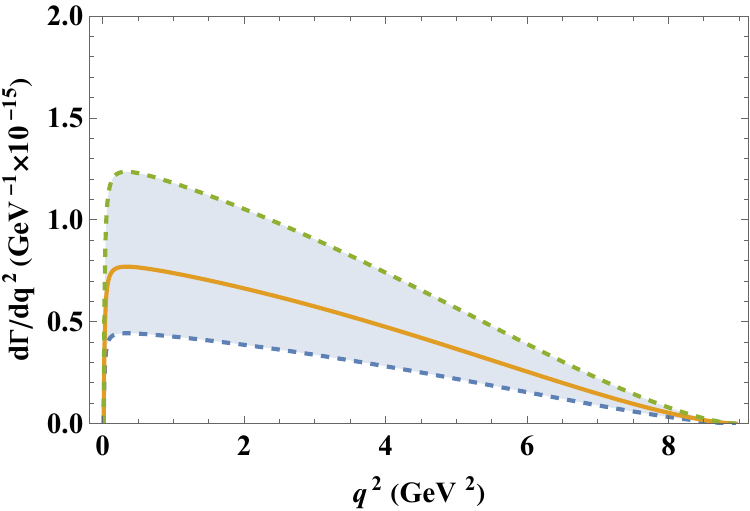}
	 \includegraphics[width=0.3\textwidth]{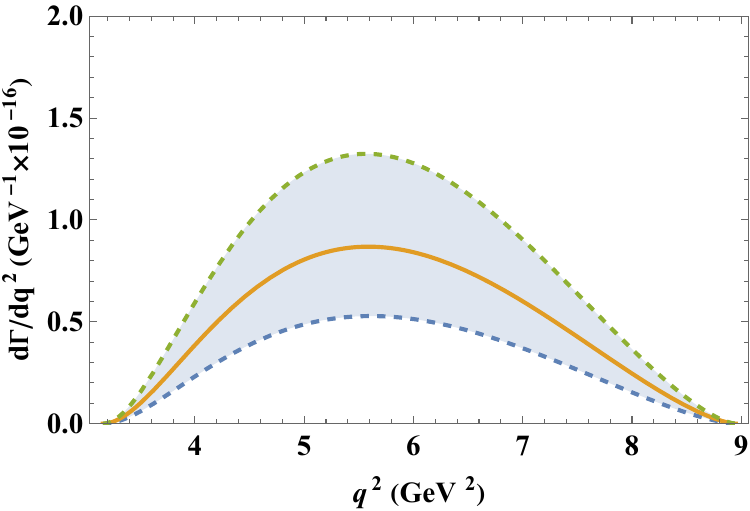}
	 \caption{Differential decay width of $\Lambda_b^0\to\Lambda_c(2625)^+\ell^-\overline\nu_\ell$ with $\ell=e$ (left), $\mu$ (middle) and $\tau$ (right).} \label{decay width 2625}
	 \end{figure}

   In order to obtain the decay widths and branching fractions of $\Lambda_b^0 \to \Lambda_c(2860)^+/\Lambda_c(2625)^+\ell^-\overline{\nu}_\ell$, we will utilize the relations between helicity amplitudes and weak decay form factors. The formula for differential decay width of semileptonic decay, with the inclusion of lepton mass, can be written as referenced in \cite{Faessler:2009xn,Gutsche:2017wag}:  
   
     Based on these calculational procedures and input parameters, along with the Fermi constant $G_F = 1.166 \times 10^{-5} \text{ GeV}^{-2}$, the graphical representations of the differential decay widths for the semileptonic decay channels of $\Lambda_b^0 \to \Lambda_c(2860)^+\ell^- \overline{\nu}_\ell$ and $\Lambda_b^0\to\Lambda_c(2625)^+\ell^-\overline{\nu}_\ell$ can be obtained. As examples, we plot the differential decay widths graphics in Fig. \ref{decay width 2860} and \ref{decay width 2625}.
			
    The absolute branching fractions of the semileptonic decays $\Lambda_b^0 \to \Lambda_c(2860)^+ \ell^- \overline{\nu}_\ell$ and $\Lambda_b^0 \to \Lambda_c(2625)^+ \ell^- \overline{\nu}_\ell$ depend on the mean lifetime of the $\Lambda_b^0$ baryon and the value of the CKM matrix element $|V_{cb}|$. For this work, according to the PDG average values, the lifetime of $\Lambda_b^0$ baryon is adopted as $\tau_{\Lambda_b^0} = (1.471 \pm 0.009) \times 10^{-12} \text{ s}$. However, $|V_{cb}|$ is not a precise value up to now. Therefore, for a rough estimation, we resort to the values obtained from the fits of inclusive and exclusive semileptonic decays of the $B$ meson in the PDG, specifically $|V_{cb}| = (40.8 \pm 1.4) \times 10^{-3}$.  
	  
	\begin{table}[h]
	\centering
	\caption{Decay width and branching fractions of $\Lambda_b^0\to\Lambda_c(2625)^+/\Lambda_c(2860)^+\ell^-\nu_\ell$ and compared with others ($\times 10^{-3}$).}
	\begin{tabular}{ccccccc}\hline
	Decay channel &  This work  & LCSR \cite{Aliev:2023tpk}  & LFQM\cite{Li:2022hcn} &CCQM\cite{Gutsche:2018nks} &HQSS\cite{Nieves:2019kdh}&  Exp.\cite{ParticleDataGroup:2024cfk} \\ \hline
	$\Lambda_b^0\to\Lambda_c(2625)^+e^-\nu_e$ & $8.25\pm0.22$ & $14.4\pm5.6$     &$16.53\pm1.14$&$1.7\pm0.3$&- & $13^{+6}_{-5}$ \\
	$\Lambda_b^0\to\Lambda_c(2625)^+\mu^-\nu_e$ & $8.12\pm0.22$  & $14.2\pm5.6$    &$16.41\pm1.13$&$1.7\pm0.3$&$35^{+13}_{-12}$& $13^{+6}_{-5}$ \\
	$\Lambda_b^0\to\Lambda_c(2625)^+\tau^-\nu_\tau$ & $0.642\pm0.016$  &  $1.1\pm0.4$   &$1.688\pm0.116$&$0.18\pm0.04$&$3.8^{+0.9}_{-0.8}$&-  \\
	$\Lambda_b^0\to\Lambda_c(2860)^+e^-\nu_e$ & $4.32\pm0.01$ &  - &  -&-&-&-  \\
	$\Lambda_b^0\to\Lambda_c(2860)^+\mu^-\nu_e$ & $4.18\pm0.01$  & -  &- &- &-&-  \\
	$\Lambda_b^0\to\Lambda_c(2860)^+\tau^-\nu_\tau$ & $0.0975\pm0.0001$  &- & -  &- &-&-  \\ \hline
	\end{tabular} \label{br}
	\end{table}
	 	 
	 Taking these input parameters into account, we derived the decay widths of the decay channel $\Lambda_b^0 \to \Lambda_c(2860)^+\ell^-\bar\nu_\ell$ and $\Lambda_b^0 \to \Lambda_c(2860)^+\ell^-\bar\nu_\ell$ and calculated their absolute branching fractions. These results and compared with other theoretical such as QCD light-cone sum rule (LCSR), light-front quark model (LFQM), covariant confined quark model (CCQM) and heavy quark spin symmetry (HQSS) are tabulated in Table \ref{br}.  
	
    \section{Conclusion} \label{sec:V}

Within the framework of QCD light-cone sum rules, we calculate the form factors for the transitions $\Lambda_b^0\to\Lambda_c(2860)^+$ and $\Lambda_b^0\to\Lambda_c(2625)^+$. The obtained form factors are then extrapolated to the physical region of the semileptonic decays $\Lambda_b^0\to\Lambda_c(2860)^+\ell^-\nu_\ell$ and $\Lambda_b^0\to\Lambda_c(2625)^+\ell^-\nu_\ell$, and we compute their differential decay widths and branching fractions, as shown in Table \ref{br}.

In the calculation of the form factors using QCD light-cone sum rules, we take into account both positive- and negative-parity states of the final-state spin-3/2 baryon, separate the Lorentz structures associated with the final-state spin-1/2 baryon, and consider only those of the spin-3/2 state. Our calculated branching fraction for the semileptonic decay $\Lambda_b^0\to\Lambda_c(2625)^+\ell^-\overline{\nu}_\ell$ is found to be in good agreement with experimental measurements and other theoretical calculations. Furthermore, we present theoretical predictions for the branching fraction of $\Lambda_b^0\to\Lambda_c(2860)^+\ell^-\overline{\nu}_\ell$, which can serve as a theoretical reference for experimental searches and analyses of this decay channel.	
	\section*{Acknowledgments}
	This work was supported by the Natural Science Foundation of Henan Province under Grant No. 262300420314, Postdoctoral Fellowship Program of CPSF under Contract No. GZC20230738, National Natural Science Foundation of China under Grants No. 12275067,  National Natural Science Foundation of China under Grants No. 12575098,  Science and Technology R\&D Program Joint Fund Project of Henan Province (Grant No.
225200810030), and Science and Technology Innovation Leading Talent Support Program of Henan Province (Grant
No. 254000510039).


   \bibliographystyle{unsrt}
   
   \bibliography{ref}
  
\end{document}